# Surface band segregation and internal convection in rotating sphere densely filled with granular material: Experiments

Weitao Sun (孙卫涛)[1,2]

[1]School of Aerospace Engineering, Tsinghua University, Beijing, 100084, China
[2]Zhou Pei-Yuan Center for Applied Mathematics, Tsinghua University, Beijing, 100084, China
Corresponding author: Weitao Sun, sunwt@tsinghua.edu.cn

**ABSTRACT**

While granular segregation in partially filled containers has been studied extensively, granular dynamics in densely filled spheres is not fully understood. Here, surface band segregation and granular convection are reported in a rotating sphere of highly compacted glass beads. Distinct from Rayleigh–Bénard convection, granular convection has a butterfly-shaped structure with vortexes of alternating layers of small/large beads, which is stable and independent of the sphere size. Two concentric interfaces at the zero tangential/norm flux are discovered, which divide the sphere into three layers from the surface to the core. The law that governs the jamming dynamics in rotating spheres remains an open question.

## I. INTRODUCTION

Granular matter has been thought of as the second most ubiquitous substance in the world following water. Unlike fluids and solids, granular matter shows rich interesting behaviors, including size segregation and convection.[1-4] Segregation occurs under periodic perturbations when grain mixtures have different sizes or densities. The large grains move to the surface under vertical shaking, which is known as the 'Brazil nut' effect.[5, 6] Despite granular matter playing substantial roles in daily life and industrial/scientific fields, an understanding of their segregation processes is still limited.

In addition to segregation, granular convection has been known for years.[7-11] Loosely compact granular systems under external excitation show instabilities similar to fluids.[12] Although granular convection and segregation have been extensively studied both in experiments and computational models,[7, 8, 11, 13-28] little attention has been given to the dynamics of granular mixtures in densely filled rotating spheres. Dense granular systems show complex behaviors such as those seen in the glass transition of supercooled liquids,[29-31] which are far from well understood. Recent experiments begin to uncover segregation patterns in densely filled drums and rectangular containers.[7, 9, 32, 33] Rietz and Stannarius[9] observed convection rolls combined with segregation in a nearly compactly filled flat rectangle cell. Compared with rectangular cells, granular convection in rotating spheres of dense material is of interest to geophysicists and astrophysicists[34-38] but remains poorly characterized.

This letter reports surface band segregation and convection in a rotating sphere that is tightly filled with two sizes of glass beads. The experiments reveal large bead bands that appear on the sphere surface as a result of internal granular convection. The convection shares a similar appearance with thermal convection in fluid spheres. However, granular convection has an unexpected butterfly-shaped structure with four rotating vortexes, which is different from the Rayleigh–Bénard instability. Alternating layers of large and small beads appear from the core to the periphery of the vortex. Two interfaces are found inside the sphere, one with a zero tangential component and the other with a zero normal component. The interfaces are at depths of 12% and 68% of the radius of the sphere, respectively, which divides it into three layers from the surface to the core. Although convection patterns in a heated fluid sphere with similar structures have been theoretically predicted,[1,2] direct observations of the butterfly-shaped convective structure and circulation/flux separation interfaces in a densely-filled rotating sphere have not been reported.

## II. EXPERIMENTAL SETUP

A 20 cm diameter transparent sphere is clamped between two arms. The left arm is driven by a stepper motor (**FIG. 1**), the right arm is mounted on the frame through a bearing, and the distance between the two arms is adjusted using a manual wheel. The sphere is densely filled with a mixture of large (0.8–1.0 mm) and small (0.4–0.6 mm) diameter glass beads with equal weight fractions. The sphere is tapped several times until it can hold no more beads. The total weight of the beads is approximately 6.9 kg.

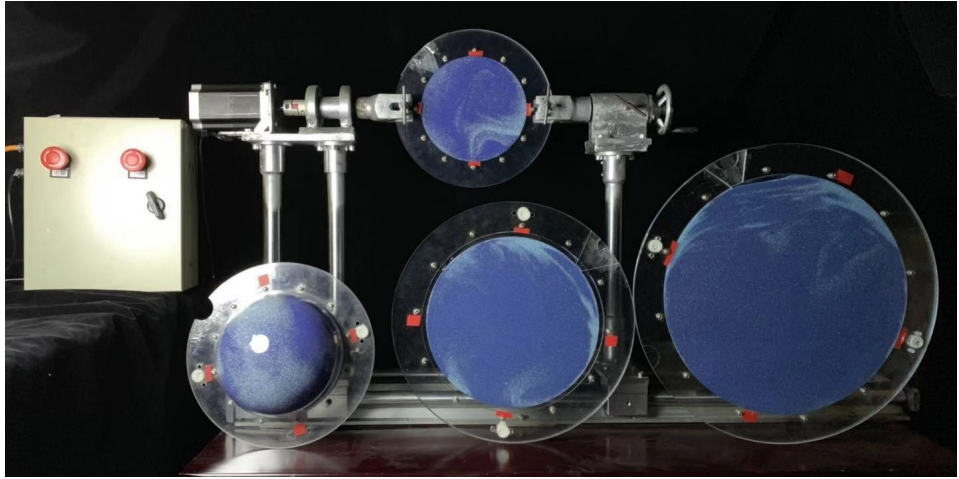

**FIG. 1.** Experimental equipment (programmable controller, rotating frame, sphere, and cells).

The sphere is driven by a motor and rotates around a horizontal axis passing through its center at a speed of 15 rounds per minute (rpm). The bead motion is recorded using a time-lapse photography (TLP) technique. The sphere stops every 15 rounds as triggered by a signal from the programmable controller. When the sphere stops rotating, the controller sends another signal through the shutter line, which triggers the camera (Nikon D7000, 6000x4000 pixels) to take a photograph. Two LED tubes are used to provide front illumination. After the photograph is captured, the controller sends a signal to the motor to rotate again until the next stop and photograph. A typical experiment will last from 2 days to a week. The photos are converted into videos to provide a continuous view of the granular flow.

## III. SURFACE BAND SEGREGATION AND INTERNAL GRANULAR CONVECTION

The TLP shows the surface band segregation and coarsening in the sphere (multimedia view). Bands with large beads (white) emerge and drift to the sphere surface. The sphere is initially filled with a mixture of large (white) and small (blue) beads [**FIG. 2**(a)]. As the rotation continues, narrow white bands parallel to the equator (great circle on the sphere perpendicular to the horizontal axis) appear and move toward the polar regions (left- and right-most parts of the sphere) [**FIG. 2**(b)]. The thin bands then begin to merge into a large band [**FIG. 2**(c)]. Finally, the white bands near the poles disappear, and the central band widens and persists until the end of the experiment after approximately 51 hours (multimedia view) [**FIG. 2**(d)].

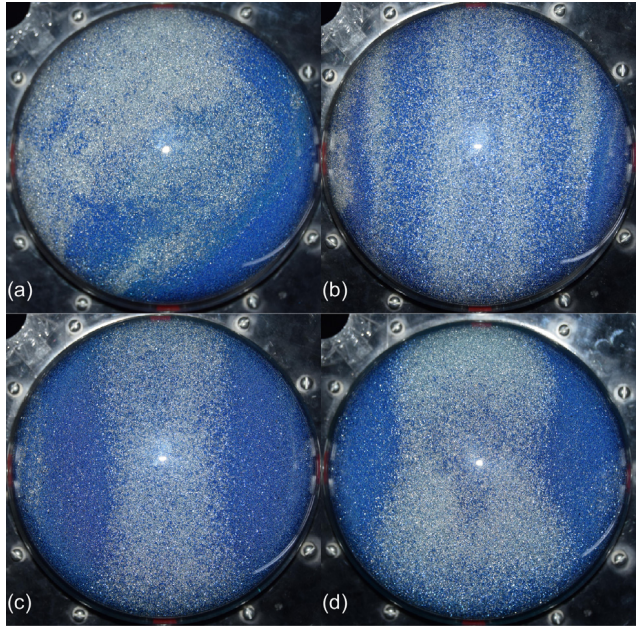

**FIG. 2.** Surface bands in a densely filled sphere at 20 cm in diameter rotating around a horizontal axis with white the large beads and blue small beads. (a)–(d) Snapshots (multimedia view) at different rotations: (a)15, (b) 2100, (c) 11700, and (d) 45750 rounds.

Larger beads segregate and form vertical bands on the sphere surface. Although axial segregation in a partially filled rotating drum was reported over half a century ago,[39, 40] the particle separation in compact-filled containers has only been recently studied.[9, 32] The tight filling prevents the beads from moving independently, which results in their slow movement throughout the sphere and is distinct from partially filled spinning spheres.[2, 21, 41] The beads inside the densely packed sphere are in close contact, and the forces between them are far more complex.

In the rotating sphere, bead movement below the surface is invisible due to the lack of transparency. To study how the bands are formed, a thin circular cell (diameter of 20 cm and thickness of 5 mm) is used as a cross-section of the sphere (**FIG. 1**). The cell consists of two closely spaced parallel circular plates with sealed edges. The cell is densely filled with a mixture of large (0.8–1.0 mm) and small (0.4–0.6 mm) diameter glass beads with equal weight fractions. **FIG. 3**(a)–3(d) show the bead movement (multimedia view) inside the cell over time (approximately 51 hours).

The volume fraction (VF) of large beads during rotation is extracted using a sliding window algorithm. A square window centered at a pixel is used to determine the VF of the large beads with a window side length of L = 9 pixels. The large beads are represented as pixels with an RGB color value within a predetermined range. The VF is obtained as $\frac{N_L}{A}$ at any position by sliding the window over a photograph of the convection and counting the number of large bead pixels ($N_L$) in the window, where $A$ is the window area. The VF is renormalized to a range [0, 31].

A previously unreported butterfly-shaped convection pattern is observed inside the cell [**FIG. 3**(e)–3(h) and **FIG. 4**(a)]. The large-bead-band (L-band) on the sphere surface is caused by the granular convection shown in **FIG. 3**(e)–3(h) (multimedia view). The formation of the convection structure consists of three stages: plume upwelling, aggregation and lateral extrusion, and subduction. In the first stage, the beads undergo a rapid process from chaotic to ordered. An upwelling plume gradually emerges from the center to the surface [**FIG. 3**(f) and Figure 3(g)]. Once the plume forms, the global convection speeds up and a conduit forms that connects the diapir with the central zone [**FIG. 3**(g)]. Both large and small beads flow through the plume channel to the top of the sphere. When the plume reaches the surface, the large beads gather near the equator [**FIG. 3**(f)].

In the second stage, as a large number of beads explosively eject at the top surface, large beads segregate and form an L-band near the equator. The newly formed L-band is pushed intermittently and moves laterally toward the poles, which causes the L-band to drift on the surface. In the third stage, the drifting L-band meets a bulk of small beads on the way to the poles. The L-band collides, dives down, and branches into two directions. One of the flow channels reverses to form a vortex below the L-band, the other continues moving toward the pole zones to form the pole band. The depth of the vortex is approximately one-third of the sphere radius. Some of the large beads flow back into the plume through the vortex, which are resupplied by an influx of large beads. This eventually forms a stable butterfly-shaped structure with beads constantly bubbling up from the top (**FIG. 3**(h) and **FIG. 4**). When the convection reaches a steady-state, the large beads form a stable grand surface band near the equator.

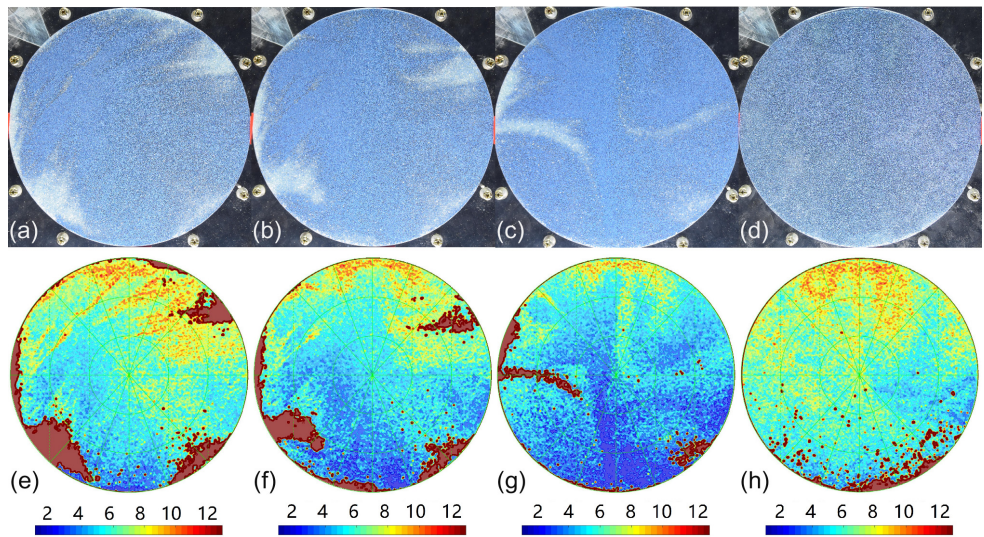

**FIG. 3.** Granular convection and volume fraction (VF) distribution in a rotating circular cell that is 20 cm in diameter where the large grains are white and small grains are blue. (a)–(d) Convection snapshots (multimedia view) and (e)–(h) VF snapshots (multimedia view) of large beads at different rounds. The VF is renormalized to a range [0, 31], and the color bar is in a range [0,12] for clear visualization. (a) and (e) 15, (b) and (f) 2100; (c) and (g) 11700; (d) and (g) 45750 rounds.

Although the large-scale convective structure remains stable, the glass beads still flow at a small scale over a long time (from days to a week). The jammed glass beads show local instabilities and form rotating vortexes. There is a strong contrast between the local flow instability and global convection/segregation stability, indicating a balance between the grain-scale dynamics and the large-scale equilibrium state. Due to the large inter-granular space ratio, the effective mass density may be lower in the large glass bead region. However, the low-density region is not made up of the same fixed batch of large glass beads. As the beads circulate through the container, large beads become more concentrated when they enter the low-density region. When leaving this region, they become more dispersed within the small glass beads. This phenomenon closely resembles density waves and explains the spiral arms of a disk nebula.[42] Density waves in granular matter have been studied over recent decades.[43-45] To our knowledge, this is the first time a butterfly-shaped density wave structure has been observed in a tightly packed rotating sphere. The granular circulation system appears as a self-perpetuating machine driven by external gravity and internal friction from the glass beads. The governing law of such complex dynamics is still missing.

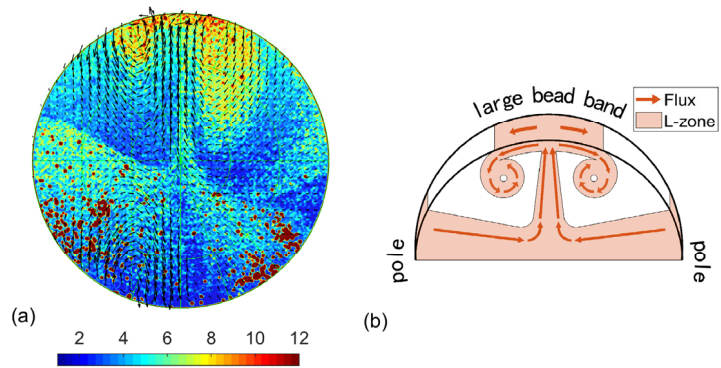

**FIG. 4.** Butterfly-shaped convection structure at 70500 rounds. (a) Large bead VF distribution and granular flow velocity field covering it. The arrows that represent the magnitude and direction of the velocity are sparsely sampled and renormalized to give a clear visualization. (b) Diagram of granular convection causing surface bands. The large-bead-zone (L-zone) has a high VF of large beads.

Jammed beads are subjected to gravitational fields. The direction of gravity acting on the beads in the upper half of the sphere is from the top surface to the center, and that on the beads in the lower half of the sphere is from the center to the bottom surface. The top beads periodically become the bottom beads during rotation. As a result, the direction of gravity on the beads also changes periodically and switches between toward and away from the center of the sphere. Under the periodically changing force direction, vortices emerge in the four quadrants of the granular convection field to form relatively independent convection regions. The beads in each sub-region flow in a whirlpool and are separated into alternate layers. The velocity field shows that the vortexes in the upper-left and lower-right corners rotate counter-clockwise, while the upper-right and lower-left vortexes rotate clockwise (**FIG. 4**(a) and multimedia view). The swirl-shaped rolls are not completely isolated but exchange glass beads through grain flow. The rolls then form a global convection system and interact with each other.

The experiment is repeated with circular cells of different diameters. **FIG. 5** shows the snapshots (multimedia view) of the granular convection and VF distribution for a cell with a diameter of 30 cm, and **FIG. 6** shows the same for a cell with a diameter of 40 cm. The same convection and segregation patterns are observed in all experiments. The observations suggest that the butterfly-shaped segregation/convection structure is stable and independent of the container size. For long-term evolutions, the system eventually reaches a "quasi-stationary" structure that lasts for days.

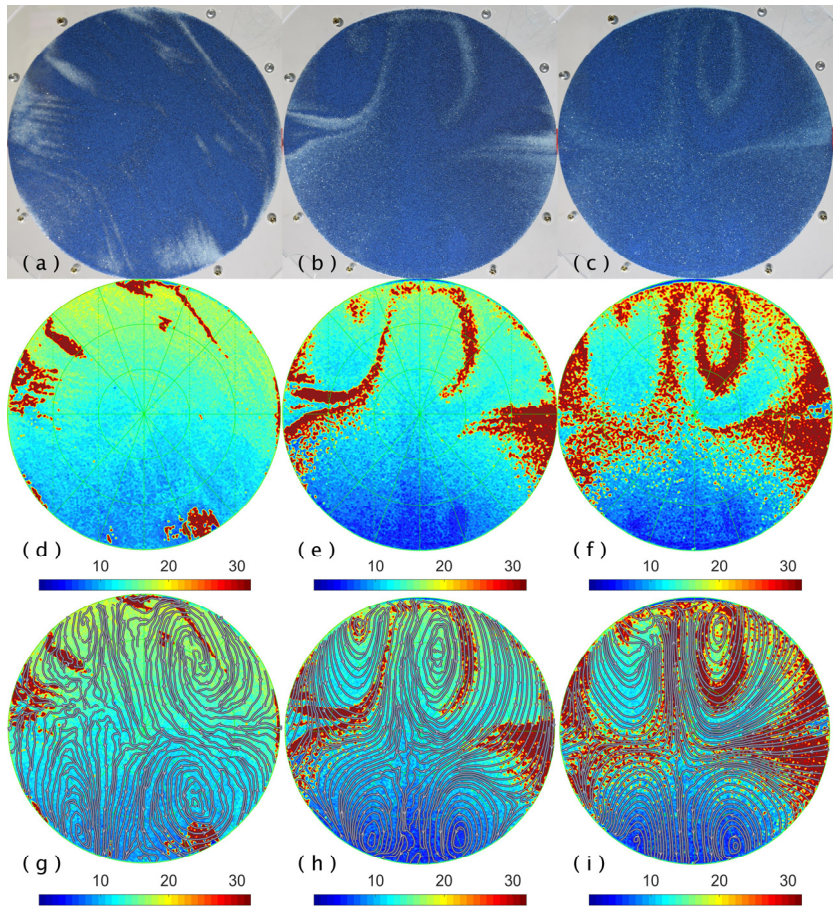

**FIG. 5.** Granular convection in a rotating circular cell with a 30 cm diameter where large grains are white and small grains are blue. (a)–(c) Convection snapshots (multimedia view) and (d)–(f) VF distribution snapshots of large beads at different rounds. The VF is renormalized to a range [0, 31]. (g)–(h) The VF snapshots with velocity field streamlines. (a), (d), and (g) 15; (b), (e), and (g) 11 700; and (c), (f), and (i) 45 750 rounds.

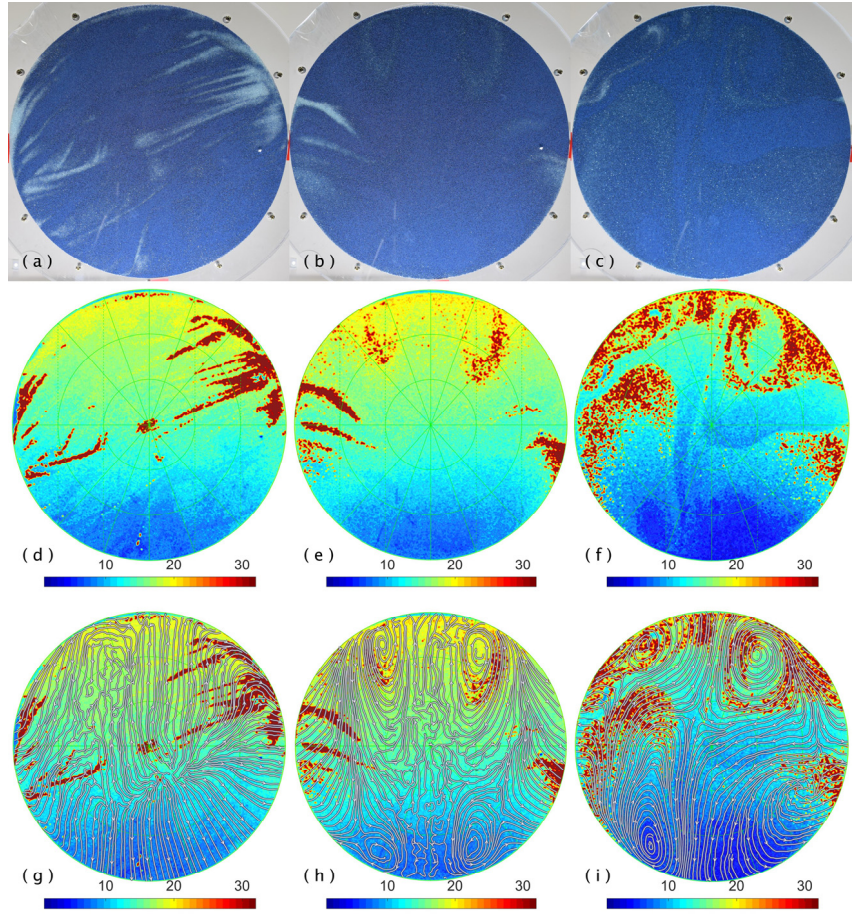

**FIG. 6.** Granular convection in a rotating circular cell 40 cm in diameter. (a)–(c) Convection snapshots (multimedia view) and (d)–(f) VF distribution snapshots of large beads at different rounds. The VF is renormalized to a range [0, 31]. (g)–(h) The VF snapshots with velocity field streamlines. (a),(d), and (g) 15; (b),(e), and (g)11700; and (c), (f), and (i) 45750 rounds.

The segregation-convection in the experiments is different from that in partially filled spherical tumblers, such as on a rotating drum, tumbler, and rectangular cells. Although granular convection has been noted, the essential discovery in this study is that a stable butterfly-shaped convection system with four rotating vortexes is observed in a densely filled sphere, rather than solid-body rotation in a half-filled tumbler.[46, 47] Experiments with cells of different sizes showed that the convection/separation pattern can be consistently repeated (multimedia view), which suggests a fundamental physics that dominates the dynamics in packed granular spheres.

## IV. INTERFACES OF ZERO-CIRCULATION AND ZERO-FLUX

The velocity fields are obtained for granular convection by tracing the position differences for beads from successive photographs. Bead displacements between two images are analyzed using a particle image velocimetry (PIV) algorithm.[48, 49] The glass beads move along curved paths with unsteady speeds. Given a concentric circular interface centered at the cell (**Fig. 7**), the tangential and normal velocities are summed to measure the rotation of granular flow in the area covered by the interface (circulation) and granular flow passing through the closed circle (flux).

The circulation and flux are calculated at a series of interfaces from the center to the surface (**Fig. 7**). The results show that the circulation and flux are non-uniformly distributed along the radius. The circulation varies from positive to negative as the radius increases. There is a circulation separation interface (CSI) with zero-circulation. From the inside of the CSI to the outside, the circulation changes from $\Gamma = 0.35$ to $\Gamma = -0.22$ mm/s. The CSI divides the sphere into an inner positive (counter-clockwise rotation) layer and outer negative layer. Outside the CSI, the granular flow is in the clockwise direction.

The flux also varies from negative to positive as the radius increases, and a flux separation interface (FSI) of zero-flux is present. The FSI divides the sphere into inner negative and outer positive layers (flux leaves a closed surface). Inside (outside) the FSI, beads flow toward the center (surface) of the sphere. The granular flux changes from $J = -0.04$ to $J = 0.01$ mm/s at the interfaces adjacent to the FSI from the inside to the outside (**Fig. 7**). The FSI is a boundary with a balanced incoming/outcoming flux.

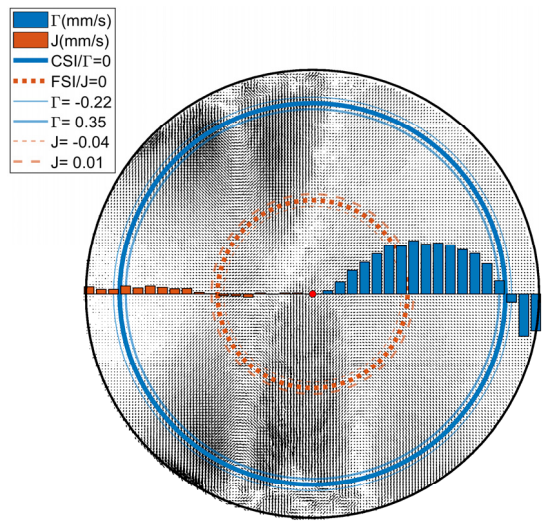

**Fig. 7.** Calculated circulation ($\Gamma$) and flux (J) of granular convection using the velocity field (background arrows) at 70 500 rotations. The bars to the left and right of center are the flux and circulation, respectively, at the interfaces of the corresponding radii.

The depths of the CSI and FSI vary with the rotation and converge to relatively stable values (**Fig. 8**). The CSI and FSI are always observed in repeated experiments with different cell diameters. Their depths appear to be independent of the sphere size, suggesting that their emergence is a fundamental phenomenon related to granular convection in rotating spheres.

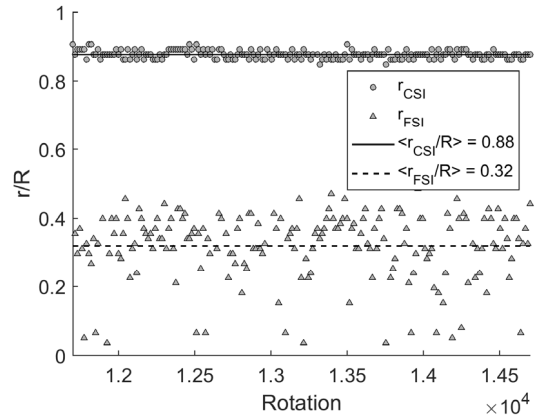

**Fig. 8.** Normalized radii (divided by sphere radius) of CSI and FSI at different rotations in a cell with a 20 cm diameter. The CSI radius converges to a stable mean of approximately 88% of the sphere radius (depth of 12% of sphere radius). The FSI radius has a stable mean of approximately 32% of the sphere radius (depth of 68% of sphere radius).

The circulation and flux separation interfaces arise naturally and divide the granular sphere into three layers from the surface to the core (**FIG. 9**). The minimum and maximum FSI depths observed in the experiment are 52% and 82% of the sphere radius, respectively. If there exists an FSI in a rotating sphere with a radius of 6371 km (approximately the radius of the earth), the depths will cover a region between 3313–5224 km. Of note, this roughly overlaps with the outer core of the Earth at approximately 2885–5115 km.[50] The CSI has an average depth of 12% of the sphere radius with relatively small deviations. If a CSI exists in a rotating sphere the size of the Earth, it would have a depth of 765 km, which is close to the upper mantle boundary at approximately 700 km.[51]

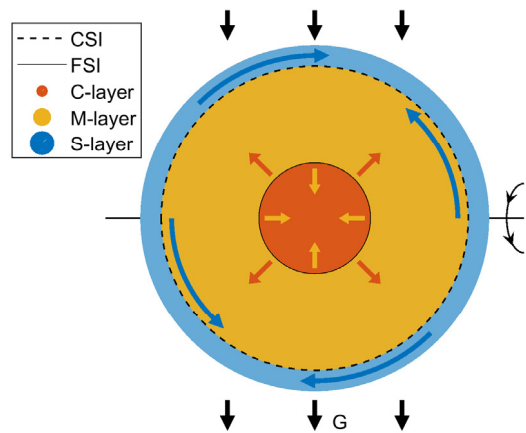

**FIG. 9.** The CSI and FSI divide the sphere into three layers: shell-layer (S-layer), middle-layer (M-layer), and core-layer (C-layer).

The evidence indicates that the internal interfaces invoke an image of Earth's internal structures. The formation of the Earth's structure, including its crust, mantle, and core, is more complex than the formation of the CSI/FSI in a rotating granular sphere. However, the analogy strongly suggests the question of whether the CSI and FSI interfaces arise spontaneously in a rotating sphere the size of the Earth. As the CSI and FSI appear independently of the sphere size, it is possible that these also form in spheres of tight granular matter approximately the size of a planet, given that the sphere is subjected to gravity similar to that in the experiment. As the internal layers of the Earth refer to those separated by seismic discontinuities, another interesting question is whether the interface of the zero-circulation and zero-flux in the granular flow field cause wave velocity discontinuities, which has yet to be answered. The physical phenomena at different scales can be understood through mathematical analyses of dimensionless models, but little work has been done to date.

## V. CONCLUSIONS

Granular bands segregate and coarsen on the surface of a rotating sphere densely filled with glass beads due to internal convection and segregation. A stable butterfly-shaped convection structure experiences three stages: plume upwelling, aggregation and lateral extrusion, and subduction. Circulation and flux separation interfaces are discovered in rotating spheres for the first time, which divides the sphere into three layers from the surface to the core. Repeated experiments show that both the convection structure and interface stratification are independent of the container size. There are still many remaining questions about how mechanical energy is transferred from external gravity to internal granular convection and surface band segregation and drift. A general theory to explain the balance between the global segregation-convection structure and local grain transport process remains an open question.

## ACKNOWLEDGMENTS

The research was supported by the National Natural Science Foundation of China (Grant no. 41874137, Grant no. 42074144) and CC Lin Special Fund.The research was supported by the National Natural Science Foundation of China (Grant no. 41874137, Grant no. 42074144) and CC Lin Special Fund.

## AUTHOR'S CONTRIBUTIONS

All authors contributed equally to this work.

## AUTHOR'S DECLARATIONS

The authors have no conflicts to disclose.

## DATA AVAILABILITY

The data that support the findings of this study are available from the corresponding author upon reasonable request.